\documentclass[12pt,oneside]{article}
\begin{document}
%\frontmatter

%\topmargin=-.35in
%\textheight=8.60in
%\oddsidemargin=0.0in
%\textwidth=6.3in

%\begin{titlepage}
\begin{center}
{\large\bf Natural TeV-Scale Gravity and coupling constant
unification, in Heterotic M-Theory, with the usual hidden and
visible sectors swapped\\}
\vspace{0.14cm}
\vspace*{.05in}
{Chris Austin\footnote{Email: chrisaustin@ukonline.co.uk}\\
%\vspace*{.3in}
\small 33 Collins Terrace, Maryport, Cumbria CA15 8DL, England\\
}
\end{center}
%\vspace*{0.8in}
\begin{center}
{\bf Abstract}
\end{center}
\noindent I consider a class of Grand Unified models, in which 
$\mathrm{E}8$ is
broken to $\mathrm{SU}(3) \times \mathrm{SU}(2) \times$ 
$\mathrm{SU}(2) \times \mathrm{SU}(2) \times \mathrm{U}(1)_Y$, then 
to $\mathrm{SU}(3) \times \mathrm{SU}(2)_{diag} \times 
\mathrm{U}(1)_Y$.  The 
breaking of $(\mathrm{SU}(2))^3$ to 
$\mathrm{SU}(2)_{diag}$ reduces the $\mathrm{SU}(2)$ coupling
constant, at unification, by a factor of $1/\sqrt{3}$, so that the 
ratio of the
$\mathrm{SU}(3)$ and $\mathrm{SU}(2)_{diag}$ coupling constants, at
unification, is equal to the ratio observed at about 1 TeV.  By 
choosing a suitable alignment of $\mathrm{U}(1)_Y$, and introducing
a generalization of the CKM matrix, the
$\mathrm{U}(1)_Y$ coupling constants of the observed fermions, at
unification, can also be arranged to have the ratios, to the 
$\mathrm{SU}(3)$ coupling constant, that are observed at about
1 TeV.  This 
suggests a model of Heterotic M-Theory, with a standard embedding of
the spin connection in one of the $\mathrm{E}8$s, but with the 
visible sector now having the $\mathrm{E}8$ that is unbroken at 
unification.  The universe is pictured 
as a thick pipe, where the long direction of the pipe represents 
the four extended dimensions, the circumference represents a 
compact six-manifold, and the radial direction represents the 
eleventh 
dimension.  The inner radius of the pipe is about $10^{-19}$ 
metres, and the outer radius of the pipe is about $10^{-14}$ 
metres.  We live on the inner surface of the pipe.  The low-energy 
generation structure and the high mass of the top quark follow from
the breaking pattern, and gravity and the Yang-Mills interactions 
are unified at about 1 TeV.  Parity breaking must
occur spontaneously in four dimensions, rather than being inherited
from ten dimensions.  The stability of the proton might be 
correlated with the entries in the CKM matrix.

\vspace{3.7cm}

\section{Introduction}

It has been observed by Arkani-Hamed, Dimopoulos, and Dvali, (ADD), 
\cite{ADD1, ADD2} that in models with $n$ extra compact dimensions, 
if the
non-gravitational fields are confined to a zone, in the compact
dimensions, whose size is small compared to the zone in which the
gravitational fields propagate, the fundamental scale of gravity can
be reduced from the Planck mass to the TeV scale, because the 
effective four-dimensional Planck mass, $M_{Pl}$, at distances 
larger than the size, $R_{ADD}$, of the gravity-only compact 
dimensions, is 
related to the fundamental gravity scale, $M_{Gr}$, by the relation:
\begin{equation}\label{ADD equation}
M_{Pl} = (R_{ADD}M_{Gr})^{\frac{n}{2}}M_{Gr}
\end{equation}
Perhaps the most natural such model is Heterotic M-Theory \cite{HW1,
HW2}.  In this model the universe is eleven-dimensional, and 
pictured as
the Cartesian product of the four extended dimensions, a
six-dimensional compact space, and a one-dimensional interval.  The
fields of eleven-dimensional supergravity propagate in the
eleven-dimensional bulk, and couple to a ten-dimensional
supersymmetric Yang-Mills theory, with an $\mathrm{E}8$ gauge group,
at each of the two ten-dimensional ``ends'' of the universe.  If
there is \mbox{N = 1} supersymmetry in the four extended 
dimensions, then,
as originally shown by Candelas, Horowitz, Strominger, and Witten,
(CHSW), \cite{CHSW}, the six-dimensional compact space has 
$\mathrm{SU}(3)$ holonomy, and the spin connection, on the 
six-dimensional compact space, is naturally embedded in the 
$\mathrm{E}8$ gauge group at one of the two ``ends'' of the 
universe, leaving the $\mathrm{E}8$ gauge group at the other 
``end'' of the universe, unbroken.  The different ``instanton 
numbers'' of the $\mathrm{E}8$ gauge fields, at the two 
ten-dimensional ``ends'' of the universe, then force the 
six-dimensional compact space to have a different volume at each of 
the two ``ends'' of the universe, as demonstrated by Witten,
\cite{Witten}.  The volume of the six-dimensional compact space is 
larger at the ``end'' of the universe
that has the spin connection embedded in its $\mathrm{E}8$ gauge
field, and this is usually assumed to be the visible ``end'' of the
universe, because the embedding of the spin connection breaks the
$\mathrm{E}8$ gauge group to $\mathrm{E}6$, and enables the Weyl
condition on the ten-dimensional spinor to be transmitted to a
four-dimensional spinor in the \textbf{27} of $\mathrm{E}6$, thus
giving a natural explanation for the observed parity violation of
the couplings of the fermions to the $\mathrm{SU}(2)$ gauge field.  

It is possible for the volume of the six-dimensional compact
space, at the ``end'' of the universe where it is larger, to be 
much larger than at the ``end'' of the universe where it is smaller,
and Witten, \cite{Witten}, considered such a limit, in order to
derive a lower bound on Newton's constant, assuming that the 
visible ``end'' of the universe is the ``end'' where the volume of 
the six-dimensional compact space is larger, and showed that this 
bound is approximately saturated, in the framework of conventional 
Grand Unification.  Witten effectively worked to first order in the 
dimensionless parameter $\frac{1}{m_{11}^3\sqrt{V}}$, where $m_{11}$
is the eleven-dimensional Planck mass, and $V$ is the proper
volume of the six-dimensional compact space, at the ``end'' of the 
universe where it is largest.  This parameter has to
be small compared to 1, for the eleven-dimensional description to be
valid.  Witten showed that, within his linearized approximation, the
proper volume of the six-dimensional compact space decreased 
linearly with proper distance, along the eleventh dimension, from 
the ``end'' where the volume is largest, with the coefficient being 
a constant factor of order 1, times
\begin{equation}\label{volume decrease coefficient}
\frac{1}{m_{11}^3}\left\vert\int_X\omega\wedge\frac{\mathrm{tr}F
\wedge F - \frac{1}{2}\mathrm{tr}R\wedge R}{8\pi^2}\right\vert
\end{equation}
Here $\omega$ is the Kahler form of the six-dimensional compact 
space $X$, and $F$ is the field strength of either of the 
$\mathrm{E}8$'s.  The integral in
(\ref{volume decrease coefficient}) is equal to 
$V^{\frac{1}{3}}$, times a constant factor of order 1, that is the 
same for all models, times a number $M$, that depends on 
integer-valued topological invariants, and has a fixed value for a
particular model.  Thus, up to a constant 
factor of order 1, the proper length, $\rho$, of the universe, in 
the eleventh dimension, is limited by
\begin{equation}\label{limit on rho}
\rho < \frac{m_{11}^3V^{\frac{2}{3}}}{\vert M\vert}
\end{equation}
since the volume of the six-dimensional compact space vanishes, at
the end of the universe where it is smaller, when this bound is
saturated.

$M$ has usually been of order 1, in specific models.  Here I want to
consider, instead, models in which $\vert M\vert$ has a
fixed \emph{large} value.  It is hard to find a formula for $M$ in 
the literature, and it is by no means clear that models, with 
$\vert M\vert$ large, actually exist.  Comparing two papers 
\cite{Choi1, Choi2} by Choi et al., suggests that $M$ might have the form 
of a sum of integer-valued quantities, over the independent periods 
$\omega_I$ of the integer (1,1) cohomology, divided by the cube root
of a triple sum of other integer-valued quantities, over the 
$\omega_I$.  I am not certain that this is the correct way to 
compare the two papers, but if it is, then since the number of
independent $\omega_I$ has the finite value $h = 
\mathrm{dim}H^{1,1}(X)$, we would expect that in general $M$ is of 
order 
$h/(h^3)^{\frac{1}{3}}$, and thus of order 1.  However, since the
integer-valued quantities in the numerator and the denominator have
different structures, it is reasonable to conjecture that models 
with large $\vert M\vert$ can nevertheless exist.  Here I shall 
simply assume that models with large $\vert M\vert$ exist, and
study their properties.

We now have two small parameters,
$\frac{1}{m_{11}^3\sqrt{V}}$ and $\frac{1}{\vert M\vert}$.  However,
$\vert M\vert$ cannot be large compared to $m_{11}^3\sqrt{V}$, 
since, with (\ref{limit on rho}), that would imply $\rho\ll 
V^{\frac{1}{6}}$ even when the
inequality in (\ref{limit on rho}) is saturated, which is a 
geometric 
impossibility, since the smallest possible proper volume of the
six-dimensional compact space, at a proper distance $\rho$ from the
end of the universe where it is largest, is about $(V^{\frac{1}{6}} 
- \rho)^6$, which for $\rho\ll V^{\frac{1}{6}}$ is not much smaller
than $V$, whereas we already know that when (\ref{limit on rho}) is 
saturated, the volume of the six-dimensional compact space
vanishes, at the end of the universe where it is smaller.  The 
largest possible value of $\vert M\vert$ is about
$m_{11}^3\sqrt{V}$, so that for fixed $M$, the smallest possible
value of $V$ is about $\frac{M^2}{m_{11}^6}$.  If $V$ takes its
smallest possible value, for a given fixed value of $M$, then when
the inequality in (\ref{limit on rho}) is saturated, we have 
$\rho\simeq
V^{\frac{1}{6}}$, so that the seven compact dimensions, which, in a
pictorial two-dimensional representation, looked like a cylinder to 
start with, have changed shape, first into a truncated cone, and 
finally into a disk with a hole in the middle.  When $V$ takes this 
minimum value, for the given fixed value of $M$, the upper bound on 
$\rho$ given by (\ref{limit on rho}), and the lower bound on $\rho$ 
given by the geometrical constraint, precisely coincide.  Such a 
configuration has a smaller range of allowed small deformations 
than a more general configuration, so it is reasonable to 
conjecture that such a configuration might be stable.  This gives a 
natural realization of the ADD proposal, if we assume that the 
visible ``end'' of the universe is actually the one where the 
volume of the six-dimensional compact space is smaller, which is the
one whose $\mathrm{E}8$ gauge group is \emph{not} broken by 
embedding the spin connection into it.  If the six-dimensional
compact space is roughly isotropic, we can use equation (\ref{ADD equation}),
with $n = 7$.  For example, if $M_{Gr}$ is 
about 1 TeV, $R_{ADD}M_{Gr}$ is about 40,000, and $R_{ADD}$ is 
about $10^{-14}$ metres.  Identifying $M_{Gr}$ with $M_{11}$, we see
that the fixed instanton number, $M$, needed to force $R_{ADD}$ up 
to this value, is about $10^{14}$.  This number determines the 
four-dimensional Planck mass, but is more in the nature of an
``accidental'' constant of nature, like the cosmological baryon to
entropy ratio, frozen at its present value in the first moments of
the universe.  In a pictorial three-dimensional representation, the
universe looks like a thick pipe, where the long direction 
of the pipe represents the four extended dimensions, the 
circumferential direction represents the 
compact six-manifold, and the radial direction represents the 
eleventh 
dimension.  If unification occurs at about a TeV, the inner radius 
of the pipe is about $10^{-19}$ 
metres, and the outer radius of the pipe is about $10^{-14}$ 
metres.  We live on the inner surface of the pipe.  It is important 
to bear in mind that if models with large $\vert M\vert$ do exist, 
their compact six-dimensional spaces might be extremely complicated,
a fact which this pictorial representation omits.

If unification occurs at around a TeV, the original motivation for
the CHSW construction \cite{CHSW}, which was to have $N = 1$ 
supersymmetry at low energy, in order to stabilize the gauge 
hierarchy of conventional Grand Unification, is no longer relevant.
However many of the results of CHSW are still very useful.
For example, they find that if the four extended dimensions are
maximally symmetric, then they must be flat, so that the 
four-dimensional cosmological constant vanishes.  Furthermore 
Witten's M-Theory calculation \cite{Witten}, which I used above, is
based on finding a solution of the eleven-dimensional field 
equations, that preserves supersymmetry.  Therefore I shall assume
that all the assumptions made by CHSW and Witten still apply, with
the single exception that the assumed hidden and visible sectors are
swapped.

Choosing the visible end of the universe to be the one without the
spin connection embedded in its gauge group, means that the CHSW
mechanism, for generating a chiral spinor in four dimensions, from 
the Weyl condition on the spinor in ten dimensions, no longer
operates.  The four-dimensional action, obtained by dimensional
reduction, will be invariant under parity, so that parity invariance
must be spontaneously broken in four dimensions.  On dimensional 
reduction, the ten-dimensional Yang-Mills spinor field reduces to a 
sum of terms, each of which has the form of a Cartesian product of 
a four-dimensional spinor field, depending on position in the four 
extended dimensions, and a six-dimensional spinor field, depending 
on position in the six-dimensional compact space.  There is one 
term in the sum for each independent normal mode of the spinor 
field on the six-dimensional compact space, and each term in the 
sum corresponds to an independent four-dimensional spinor field.

CHSW find that when the holonomy group of the six-dimensional 
compact space is precisely $\mathrm{SU}(3)$, which is one of the 
conditions to have precisely $N = 1$ supersymmetry in four 
dimensions, there are precisely two real covariantly constant 
spinors, $\eta$ and $i\gamma\eta$, on the six-dimensional compact 
space.  It is natural to assume that $\frac{1}{2}(\eta + 
i\gamma\eta)$ and $\frac{1}{2}(\eta - i\gamma\eta)$ correspond, 
respectively, to a single massless left-handed spinor field, and a 
single massless right-handed spinor field, in four dimensions, each 
in the \textbf{248} of $\mathrm{E}8$, which is both the adjoint and 
the fundamental.  We shall find that the left-handed states, of all 
the observed fermions and antifermions, can be fitted into a single
left-handed $\mathrm{E}8$ fundamental, except for the left-handed
components of the top antiquark.  It is therefore natural to suppose
that there are no other massless fermion modes on the 
six-dimensional compact space, and that the left-handed components 
of the top antiquark come from the lightest massive fermion mode on 
the internal compact space, which is why the top quark is so heavy.

The general prospects for obtaining dynamical symmetry breaking
in $\mathrm{E}8$ Grand Unification, of the kind required to push up
the masses of the unobserved chiral partners of the observed 
fermions, have been discussed recently by Adler \cite{Adler}, who
also gives a useful history of $\mathrm{E}8$ Grand Unification, and
many references to relevant recent work.  Here I shall concentrate
on reconciling $\mathrm{E}8$ Grand Unification with the ADD 
proposal.

The possibility of realizing the ADD proposal in Heterotic M-Theory,
in the context of non-standard embeddings of the spin connection,
was noted by Benakli \cite{Benakli}, and by Cerde\~no and Mu\~noz
\cite{CM}.

\section{Gauge coupling unification}

If the ADD proposal is to be combined with Grand Unification, then
the gauge couplings have to unify at the TeV scale, rather than at
$10^{16}$ GeV.  One way this could work is if the running of the
coupling constants somehow accelerates, so that the couplings run to
their conventional unification values at the TeV scale, rather than 
at
$10^{16}$ GeV.  This possibility was studied by Dienes, Dudas, and
Gherghetta \cite{DDG1, DDG2}, and by Arkani-Hamed, Cohen, and Georgi
\cite{ACG}.

An alternative possibility is to embed $\mathrm{SU}(3)\times
\mathrm{SU}(2)\times\mathrm{U}(1)$ into the Grand Unification group
in an unusual way, so that the values of the coupling constants, at 
unification, are equal to their observed values, as evolved 
conventionally to the TeV scale.  Usually the coupling constant of
a simple non-Abelian subgroup of a Grand Unification group, at
unification, is equal to the coupling constant of the Grand
Unification group, irrespective of how the subgroup is embedded
in the Grand Unification group.  An exception occurs \cite{Benakli, 
ACG} if 
the initial breaking of the Grand Unification group produces $N$ 
copies of the of the simple subgroup, and the $N$ copies of the 
simple subgroup then break into their ``diagonal'' subgroup.  In 
this case, after the second stage of the breaking, the coupling 
constant of the ``diagonal'' subgroup is equal to 
$\frac{1}{\sqrt{N}}$ times the coupling constant of the Grand
Unification group.  Effectively, the gauge field, in each of the
$N$ copies of the simple non-Abelian subgroup, becomes equal to
$\frac{1}{\sqrt{N}}$ times the ``diagonal'' gauge field, plus 
massive vector terms that can be ignored at low energies.  The sum
of the $N$ copies of the Yang-Mills action, of the simple 
non-Abelian subgroup, then becomes equal to the Yang-Mills action of
the ``diagonal'' subgroup, whose coupling constant is 
$\frac{1}{\sqrt{N}}$ times the coupling constant of the Grand
Unification group.

Looking at the observed values of the reciprocals of the 
$\mathrm{SU}(3)\times\mathrm{SU}(2)\times\mathrm{U}(1)$ fine 
structure constants, at $M_Z$, normalized so as to meet at 
unification in $\mathrm{SU}(5)$ Grand Unification, \cite{GG}, 
(Mohapatra \cite{Mohapatra}, page 22):
\begin{equation}\label{fine structure constant reciprocals}
\begin{array}{ccc}
\alpha_3^{-1}(M_Z) & = & 8.47\pm.22 \\
\alpha_2^{-1}(M_Z) & = & 29.61\pm.05 \\
\alpha_1^{-1}(M_Z) & = & 58.97\pm.05
\end{array}
\end{equation}
we see that they are quite close to being in the ratios 1, 3, 6.

If we evolve them in the MSSM, \cite{Mohapatra}, then 
$\alpha_3^{-1}$ and
$\alpha_2^{-1}$ reach an exact ratio of 1, 3, at 1.32 TeV, at which
point $\alpha_3^{-1}$ is equal to 9.73.  At this point, 
$\alpha_1^{-1}$ is equal to 56.20, which is 4\% off being 6 times
$\alpha_3^{-1}$.  Alternatively, if we evolve them in the SM,
\cite{Rosner}, then $\alpha_3^{-1}$ and $\alpha_2^{-1}$ reach an 
exact ratio of 1, 3, at 413 GeV, at which point $\alpha_3^{-1}$ is
equal to 10.12.  At this point, $\alpha_1^{-1}$ is equal to 58.00,
which is 4\% off being 6 times $\alpha_3^{-1}$.

Thus it is natural to consider the breaking of $\mathrm{E}8$ to
$\mathrm{SU}(3)\times\mathrm{SU}(2)\times\mathrm{SU}(2)\times
\mathrm{SU}(2)\times\mathrm{U}(1)_Y$, 
followed by the breaking of $(\mathrm{SU}(2))^3$ to
$\mathrm{SU}(2)_{diag}$, and seek an embedding of 
$\mathrm{U}(1)_Y$ that gives the correct hypercharges at 
unification.  I have summarized the required left-handed fermions of
the first generation, together with their hypercharges, $Y$, 
\cite{Rosner}, the coefficients of their $\mathrm{U}(1)_Y$ couplings
in $\mathrm{SU}(5)$ Grand Unification, and the required coefficients
of their $\mathrm{U}(1)_Y$ couplings, in Table \ref{T1}.  Here I 
have assumed that $\alpha_3^{-1}$ and $\alpha_1^{-1}$ are in the
ratio 1, 6, at unification, but it would be useful to study models 
that achieve this within a few percent, since the correct form of 
running to unification is not yet known.  Since the running of the
coupling constants is always by small amounts, the additional states
in these models, not yet observed experimentally, will not alter the
unification mass, or the value of the $\mathrm{SU}(3)$ coupling
constant at unification, which is equal to the $\mathrm{E}8$ 
coupling constant at unification, by a large amount.  Thus this 
class of models generically predicts that the unification mass is
about a TeV, and the $\mathrm{E}8$ fine structure constant, at 
unification, is about $\frac{1}{10}$.

\begin{table}
\begin{center}
\begin{tabular}{|c|c|c|c|c|}\hline
\multicolumn{5}{|c|}{First generation LH states} \\ \hline
Multiplet & Y &
$\begin{array}{c}\mathrm{SU}(3)\times\mathrm{SU}(2) \\
\mathrm{content} \end{array}$ &
SU(5) coefficient & $\begin{array}{c}\mathrm{required} \\
\mathrm{coefficient} \end{array}$ \\ 
\hline
$\left( \begin{array}{ccc}u_R & u_G & u_B \\
d_R & d_G & d_B \end{array} \right)$         & $\frac{1}{3}$ &
(3,2) & $\frac{1}{\sqrt{60}}$ & $\frac{1}{\sqrt{360}}$ \\ \hline
$\left( \begin{array}{ccc}\bar{u}_R & \bar{u}_G & \bar{u}_B 
 \end{array} \right)$ & $-\frac{4}{3}$ & $(\bar{3},1)$ &
$\frac{-4}{\sqrt{60}}$ & $\frac{-4}{\sqrt{360}}$ \\ \hline
$\left( \begin{array}{ccc}\bar{d}_R & \bar{d}_G & \bar{d}_B 
 \end{array} \right)$ & $\frac{2}{3}$ & $(\bar{3},1)$ &
$\frac{2}{\sqrt{60}}$ & $\frac{2}{\sqrt{360}}$ \\ \hline
$\left( \begin{array}{c}\nu_e \\ e^- \end{array} \right)$ & $-1$ &
(1,2) & $\frac{-3}{\sqrt{60}}$ & $\frac{-3}{\sqrt{360}}$ \\ \hline
$\left( \begin{array}{c} e^+ \end{array} \right)$ & $2$ &
(1,1) & $\frac{6}{\sqrt{60}}$ & $\frac{6}{\sqrt{360}}$ \\ \hline
$\left( \begin{array}{c} \bar{\nu}_e \end{array} \right)$ & $0$ &
(1,1) & absent & $0$ \\ \hline
\end{tabular}
\caption{\label{T1}
Weak hypercharge, $\mathrm{SU}(3)\times\mathrm{SU}(2)$ assignments,
coefficient of the coupling to the $\mathrm{U}(1)_Y$ vector boson in
$\mathrm{SU}(5)$, and the required coefficient of the coupling to
the $\mathrm{U}(1)_Y$ vector boson, for the left-handed fermions of the first
generation.}
\end{center}
\end{table}

Witten \cite{Witten} finds that the Grand Unification fine structure
constant, at unification, is equal to $\frac{(4\pi)^{\frac{2}{3}}}
{2m_{11}^6V_v}=\frac{2.7}{m_{11}^6V_v}$, where $m_{11}$ is the 
eleven-dimensional Planck mass, and $V_v$ is the volume of the 
six-dimensional compact space at the visible end of the universe.
Hence $m_{11}^6V_v=27$, so if $R_6$ denotes the diameter of the 
inner surface of the pipe, then $m_{11}R_6$ is about 1.7.  Thus 
$\frac{1}{R_6}$, the Grand Unification mass, and the 
eleven-dimensional Planck mass can all be close to 1 TeV.  Witten's
long wavelength expansion \cite{Witten} is a good approximation 
throughout most of the volume of the pipe, but breaks down near the 
inner surface of the pipe.  The Yang-Mills coupling constants of 
the fields on the outer surface of the pipe are extremely small, and
they are probably effectively free fields, even at cosmological 
distances.  However they will interact gravitationally, with the 
same value of Newton's constant as we observe, and are candidates to
form part of the dark matter of the universe.

It is convenient to use an $\mathrm{SU}(9)$ basis for $\mathrm{E}8$.
On breaking $\mathrm{E}8$ to $\mathrm{SU}(9)$, the \textbf{248} of
$\mathrm{E}8$ splits to the \textbf{80}, \textbf{84}, and 
$\mathbf{\overline{84}}$ of $\mathrm{SU}(9)$.  Here the \textbf{80} is 
the adjoint of $\mathrm{SU}(9)$, the \textbf{84} has three totally
antisymmetrized $\mathrm{SU}(9)$ fundamental subscripts, and the
$\mathbf{\overline{84}}$ has three totally antisymmetrized 
$\mathrm{SU}(9)$ antifundamental subscripts.  We can proceed in 
close analogy to the $\mathrm{SU}(5)$ model.  The 
fundamental representation generators $\left(t_\alpha\right)_{ij}$ 
are normalized to satisfy \cite{Rosner}
\begin{equation}\label{normalization}
\mathrm{tr}\left(t_\alpha t_\beta\right)=\frac{\delta_{\alpha\beta}}
{2}
\end{equation}

The generators of the required representations are as follows:
\begin{equation}\label{Antifundamental}
\textrm{\emph{Antifundamental}}\qquad\qquad\qquad
\left(T_\alpha\right)_{ij}=-\left(t_\alpha\right)_{ji}
\qquad\qquad\qquad\qquad\qquad\qquad\qquad
\end{equation}
\begin{equation}\label{Adjoint}
\textrm{\emph{Adjoint}}\qquad\qquad\qquad\:\;
\left(T_\alpha\right)_{ij,km}=\left(t_\alpha\right)_{ik}\delta_{mj}
-\delta_{ik}\left(t_\alpha\right)_{mj}
\qquad\qquad\qquad\qquad\qquad\:\;
\end{equation}
\begin{equation}\label{84}
\textrm{\textbf{84}}\qquad\qquad\qquad
\left(T_\alpha\right)_{ijk,mpq}=\frac{1}{6}\left(\left(t_\alpha
\right)_{im}\delta_{jp}\delta_{kq}\pm\textrm{seventeen terms}\right)
\qquad\qquad\qquad
\end{equation}
\begin{equation}\label{84 bar}
\mathbf{\overline{84}}\qquad\qquad\qquad
\left(T_\alpha\right)_{ijk,mpq}=\frac{1}{6}\left(-\left(t_\alpha
\right)_{mi}\delta_{pj}\delta_{qk}\pm\textrm{seventeen terms}\right)
\qquad\qquad\qquad
\end{equation}
where the additional terms in (\ref{84}) and (\ref{84 bar}) 
antisymmetrize with respect to permutations of $(i,j,k)$, and with
respect to permutations of $(m,p,q)$.  We can check directly that
these generators satisfy the same commutation relations as
$\left(t_\alpha\right)_{ij}$, with the same structure constants.

The breaking of $\mathrm{E}8$ to $\mathrm{SU}(3)\times\mathrm{SU}(2)
\times\mathrm{SU}(2)\times\mathrm{SU}(2)\times\mathrm{U}(1)_Y$ can 
be studied by analyzing the breaking of
$\mathrm{SU}(9)$ to $\mathrm{SU}(3)\times\mathrm{SU}(2)
\times\mathrm{SU}(2)\times\mathrm{SU}(2)\times\mathrm{U}(1)_Y$.  It 
is convenient to use block matrix notation.
Each $\mathrm{SU}(9)$ fundamental index is replaced by a pair of
indexes, an upper-case letter and a lower-case letter.  The 
upper-case letter runs from 1 to 4, and indicates which subgroup in
the sequence $\mathrm{SU}(3)\times\mathrm{SU}(2)
\times\mathrm{SU}(2)\times\mathrm{SU}(2)$ the block belongs to.  
Thus $B=1$ denotes the $\mathrm{SU}(3)$, $B=2$ denotes the first
$SU(2)$, $B=3$ denotes the second $SU(2)$, and $B=4$ denotes the
third $SU(2)$.  The lower-case index is a fundamental index for the
subgroup identified by the upper-case index it belongs to.  It is
important to note that the range of a lower-case index depends on 
the value of the upper-case index it belongs to, so we have to keep
track of which lower-case indexes belong to which upper-case 
indexes.  Each $\mathrm{SU}(9)$ antifundamental index is treated in 
the
same way, except that the lower-case index is now an antifundamental
index for the appropriate subgroup.  The summation convention is
applied to both upper-case letters and lower-case letters that 
derive from an $\mathrm{SU}(9)$ fundamental or antifundamental 
index, but we have to remember that lower-case indexes are to be 
summed over first, because their ranges of summation depend on the
values of the upper-case indexes they belong to.  Each 
$\mathrm{SU}(9)$ adjoint representation index, which in the notation
above, is a lower-case Greek letter, is replaced by a pair of 
indexes, an
upper-case letter and a lower-case letter, where the upper-case 
letter runs from 1 to 5, and identifies which subgroup in the 
sequence $\mathrm{SU}(3)\times\mathrm{SU}(2)
\times\mathrm{SU}(2)\times\mathrm{SU}(2)\times\mathrm{U}(1)_Y$ a 
generator belongs to, and the lower-case 
letter runs over all the
generators of the subgroup identified by the upper-case letter it
belongs to.  When an upper-case adjoint representation index takes 
the value 5, the associated lower-case index takes a single
value, 1.  The summation convention is applied to a lower-case
letter that derives from an $\mathrm{SU}(9)$ adjoint representation
index, but \emph{not} to an upper-case letter that derives from an
$\mathrm{SU}(9)$ adjoint representation index.

\begin{table}
\begin{center}
\begin{tabular}{|c|c|c|c|c|}\hline
\multicolumn{5}{|c|}{States in the \textbf{80}} \\ \hline
Blocks & $\begin{array}{c}\textrm{Number of} \\
\mathrm{distinct} \\ \mathrm{blocks} \end{array}$ & 
$\begin{array}{c}\mathrm{SU}(3)\times\mathrm{SU}(2) \\
\mathrm{content} \end{array}$ &
$\begin{array}{c}\textrm{Number of} \\
\mathrm{states} \end{array}$ & 
$\begin{array}{c}\textrm{coefficient} \\
\textrm{of coupling} \\ \textrm{to U(1)} \end{array}$ \\ 
\hline
$\psi_{11}$ & 1 & (8,1) & 8 & 0 \\ \hline
$\psi_{22}$ & 1 & (1,3) & 3 & 0 \\ 
$\psi_{33}$ & 1 & (1,3) & 3 & 0 \\ 
$\psi_{44}$ & 1 & (1,3) & 3 & 0 \\ \hline
$\begin{array}{c}\psi_{\mathrm{diag}} \\ \psi_{\mathrm{diag}} \\ 
\psi_{\mathrm{diag}} \end{array}$ & 
$\begin{array}{c}\textrm{not} \\ \textrm{applicable} \end{array}$ & 
$\begin{array}{c}(1,1)+(1,1)+ \\ +(1,1) \end{array}$ & 
3 & 0 \\  \hline
$\psi_{12}$ & 1 & (3,2) & 6 & 
$\frac{\sigma_1-\sigma_2}{\theta}$ \\ 
$\psi_{13}$ & 1 & (3,2) & 6 & 
$\frac{\sigma_1-\sigma_3}{\theta}$ \\ 
$\psi_{14}$ & 1 & (3,2) & 6 & 
$\frac{\sigma_1-\sigma_4}{\theta}$ \\ \hline
$\psi_{21}$ & 1 & $(\bar{3},2)$ & 6 & 
$\frac{-\sigma_1+\sigma_2}{\theta}$ \\ 
$\psi_{31}$ & 1 & $(\bar{3},2)$ & 6 & 
$\frac{-\sigma_1+\sigma_3}{\theta}$ \\ 
$\psi_{41}$ & 1 & $(\bar{3},2)$ & 6 & 
$\frac{-\sigma_1+\sigma_4}{\theta}$ \\ \hline
$\psi_{23}$ & 1 & $(1,3)+(1,1)$ & 4 & 
$\frac{\sigma_2-\sigma_3}{\theta}$ \\ 
$\psi_{24}$ & 1 & $(1,3)+(1,1)$ & 4 & 
$\frac{\sigma_2-\sigma_4}{\theta}$ \\ 
$\psi_{34}$ & 1 & $(1,3)+(1,1)$ & 4 & 
$\frac{\sigma_3-\sigma_4}{\theta}$ \\ \hline
$\psi_{32}$ & 1 & $(1,3)+(1,1)$ & 4 & 
$\frac{-\sigma_2+\sigma_3}{\theta}$ \\ 
$\psi_{42}$ & 1 & $(1,3)+(1,1)$ & 4 & 
$\frac{-\sigma_2+\sigma_4}{\theta}$ \\ 
$\psi_{43}$ & 1 & $(1,3)+(1,1)$ & 4 & 
$\frac{-\sigma_3+\sigma_4}{\theta}$ \\ \hline
\end{tabular}
\caption{\label{T2}
The states in the \textbf{80}, organized by their
$\mathrm{SU}(3)\times\mathrm{SU}(2)\times\mathrm{SU}(2)\times
\mathrm{SU}(2)$ assignments, showing their $\mathrm{SU}(3)\times
\mathrm{SU}(2)_{diag}$ content, and the coefficients of
their couplings to a $\mathrm{U}(1)$ gauge field, parametrized as in equation (\ref{Abelian generator}).}
\end{center}
\end{table}

We can now list all the blocks in the \textbf{80}, the \textbf{84}, 
and the $\mathbf{\overline{84}}$, and display their $\mathrm{SU}(3)\times
\mathrm{SU}(2)$ content.  This is displayed in Table \ref{T2} for 
the \textbf{80}, and in Table \ref{T3} for the $\mathbf{\overline{84}}$,
with all the lower-case indexes suppressed.

\begin{table}
\begin{center}
\begin{tabular}{|c|c|c|c|c|}\hline
\multicolumn{5}{|c|}{States in the $\mathbf{\overline{84}}$} \\ 
\hline
Blocks & $\begin{array}{c}\textrm{Number of} \\
\mathrm{distinct} \\ \mathrm{blocks} \end{array}$ & 
$\begin{array}{c}\mathrm{SU}(3)\times\mathrm{SU}(2) \\
\mathrm{content} \end{array}$ &
$\begin{array}{c}\textrm{Number of} \\
\mathrm{states} \end{array}$ & 
$\begin{array}{c}\textrm{coefficient} \\
\textrm{of coupling} \\ \textrm{to U(1)} \end{array}$ \\ 
\hline
$\psi_{111}$ & 1 & (1,1) & 1 & $\frac{-3\sigma_1}{\theta}$ \\ 
\hline
$\begin{array}{ccc}\psi_{211} & \psi_{121} & 
\psi_{112} \end{array}$ & 1 & $(3,2)$ &
6 & $\frac{-2\sigma_1-\sigma_2}{\theta}$ \\ 
$\begin{array}{ccc}\psi_{311} & \psi_{131} & 
\psi_{113} \end{array}$ & 1 & $(3,2)$ &
6 & $\frac{-2\sigma_1-\sigma_3}{\theta}$ \\ 
$\begin{array}{ccc}\psi_{411} & \psi_{141} & 
\psi_{114} \end{array}$ & 1 & $(3,2)$ &
6 & $\frac{-2\sigma_1-\sigma_4}{\theta}$ \\ \hline
$\begin{array}{ccc}\psi_{221} & \psi_{212} & 
\psi_{122} \end{array}$ & 1 & $(\bar{3},1)$ &
3 & $\frac{-\sigma_1-2\sigma_2}{\theta}$ \\ 
$\begin{array}{ccc}\psi_{331} & \psi_{313} & 
\psi_{133} \end{array}$ & 1 & $(\bar{3},1)$ &
3 & $\frac{-\sigma_1-2\sigma_3}{\theta}$ \\ 
$\begin{array}{ccc}\psi_{441} & \psi_{414} & 
\psi_{144} \end{array}$ & 1 & $(\bar{3},1)$ &
3 & $\frac{-\sigma_1-2\sigma_4}{\theta}$ \\ \hline
$\begin{array}{ccc}
\psi_{123} & \psi_{213} & \psi_{231} \\
\psi_{132} & \psi_{312} & \psi_{321}
\end{array}$ & 1 & $(\bar{3},3)+(\bar{3},1)$ & 12 & $\frac{-\sigma_1-\sigma_2-\sigma_3}{\theta}$ \\ \hline
$\begin{array}{ccc}
\psi_{124} & \psi_{214} & \psi_{241} \\
\psi_{142} & \psi_{412} & \psi_{421}
\end{array}$ & 1 & $(\bar{3},3)+(\bar{3},1)$ & 12 & $\frac{-\sigma_1-\sigma_2-\sigma_4}{\theta}$ \\ \hline
$\begin{array}{ccc}
\psi_{134} & \psi_{314} & \psi_{341} \\
\psi_{143} & \psi_{413} & \psi_{431}
\end{array}$ & 1 & $(\bar{3},3)+(\bar{3},1)$ & 12 & $\frac{-\sigma_1-\sigma_3-\sigma_4}{\theta}$ \\ \hline
$\begin{array}{ccc}
\psi_{222} & \psi_{333} & \psi_{444}
\end{array}$ &
\multicolumn{4}{|c|}{these three blocks are empty} \\ \hline
$\begin{array}{ccc} \psi_{223} & \psi_{232} & \psi_{322} \end{array}$ & 1 & (1,2) & 2 & $\frac{-2\sigma_2-\sigma_3}{\theta}$ \\ 
$\begin{array}{ccc} \psi_{224} & \psi_{242} & \psi_{422} \end{array}$ & 1 & (1,2) & 2 & $\frac{-2\sigma_2-\sigma_4}{\theta}$ \\ 
$\begin{array}{ccc} \psi_{332} & \psi_{323} & \psi_{233} \end{array}$ & 1 & (1,2) & 2 & $\frac{-\sigma_2-2\sigma_3}{\theta}$ \\ 
$\begin{array}{ccc} \psi_{334} & \psi_{343} & \psi_{433} \end{array}$ & 1 & (1,2) & 2 & $\frac{-2\sigma_3-\sigma_4}{\theta}$ \\ 
$\begin{array}{ccc} \psi_{442} & \psi_{424} & \psi_{244} \end{array}$ & 1 & (1,2) & 2 & $\frac{-\sigma_2-2\sigma_4}{\theta}$ \\ 
$\begin{array}{ccc} \psi_{443} & \psi_{434} & \psi_{344} \end{array}$ & 1 & (1,2) & 2 & $\frac{-\sigma_3-2\sigma_4}{\theta}$ \\ \hline
$\begin{array}{ccc}
\psi_{234} & \psi_{324} & \psi_{342} \\
\psi_{432} & \psi_{423} & \psi_{243}
\end{array}$ & 1 &
$\begin{array}{c} (1,4)+(1,2)+ \\ +(1,2)\end{array}$ & 8 & 
$\frac{-\sigma_2-\sigma_3-\sigma_4}{\theta}$ \\ \hline
\end{tabular}
\caption{\label{T3}
The states in the $\mathbf{\overline{84}}$, organized by their
$\mathrm{SU}(3)\times\mathrm{SU}(2)\times\mathrm{SU}(2)\times
\mathrm{SU}(2)$ assignments, showing their $\mathrm{SU}(3)\times
\mathrm{SU}(2)_{diag}$ content, and the coefficients of
their couplings to a $\mathrm{U}(1)$ gauge field, parametrized as in equation
(\ref{Abelian generator}).}
\end{center}
\end{table}

The $\mathrm{SU}(9)$ generators, in the $\mathrm{SU}(3)\times
\mathrm{SU}(2)\times\mathrm{SU}(2)\times\mathrm{SU}(2)\times
\mathrm{U}(1)_Y$ subgroup, may be taken as follows, in the block 
matrix notation.
\begin{equation}\label{non-Abelian generators}
\left(t_{Aa}^{(9)}\right)_{BiCj}\ \ =\ \ \delta_{AB}\delta_{AC}
\left(t_{Aa}\right)_{ij}\qquad\qquad\qquad(1\leq A\leq4)
\qquad\qquad\qquad\quad
\end{equation}
\begin{equation}\label{Abelian generator}
\begin{array}{ccl}
\left(t_{51}^{(9)}\right)_{BiCj} & = & \frac{1}{\theta}\left(
\sigma_1\delta_{1B}\delta_{1C}\delta_{ij}+
\sigma_2\delta_{2B}\delta_{2C}\delta_{ij}+
\sigma_3\delta_{3B}\delta_{3C}\delta_{ij}+
\sigma_4\delta_{4B}\delta_{4C}\delta_{ij}\right) \\
 & = & \frac{1}{\theta}
\left(\displaystyle\sum_{A=1}^4\sigma_A\delta_{AB}\delta_{AC}
\delta_{ij}\right)
\end{array}
\end{equation}

Here $\left(t_{Aa}\right)_{ij}$ denotes the fundamental 
representation generator number $a$, of non-Abelian subgroup number 
$A$, in the list above.  Thus for $A=1$, the subgroup is 
$\mathrm{SU}(3)$, $a$ runs from 1 to 8, and $i$ and $j$ each run
from 1 to 3, while for $A=2,\textrm{ 3, or 4}$, the subgroup is
$\mathrm{SU}(2)$, $a$ runs from 1 to 3, and $i$ and $j$ each run
from 1 to 2.

$\sigma_1$, $\sigma_2$, $\sigma_3$, and $\sigma_4$ are real numbers
parametrizing the embedding of the $\mathrm{U}(1)_Y$ subgroup in
$\mathrm{SU}(9)$, and thus in $\mathrm{E}8$, and $\theta$ is a
normalization factor.

In using this notation, we have to take sensible precautions, such
as grouping within brackets, to keep track of which lower-case 
indexes belong to which upper-case indexes.  In equation
(\ref{Abelian generator}), it would be wrong to ``factor out'' the
$\delta_{ij}$, because it represents a \mbox{3 by 3} matrix in one
term, and a 2 by 2 matrix in the other three terms.

The tracelessness of $\left(t_{51}^{(9)}\right)_{BiCj}$ implies:
\begin{equation}\label{tracelessness}
0 = 3\sigma_1+2\left(\sigma_2+\sigma_3+\sigma_4\right)
\end{equation}
and the normalization condition (\ref{normalization}) implies:
\begin{equation}\label{theta squared}
\theta^2 = 
6\sigma_1^2+4\left(\sigma_2^2+\sigma_3^2+\sigma_4^2\right)
\end{equation}

As an example, I consider the states in the left-handed 
$\mathbf{\overline{84}}$.  The covariant derivative is \cite{Rosner}
\begin{equation}\label{covariant derivative}
\mathcal{D}_\mu = \partial_\mu - i g A_{\mu\alpha}T_\alpha
\end{equation}
so, for unbroken $\mathrm{SU}(9)$, the massless Dirac action in 
this case is \cite{Rosner}:
\begin{displaymath}
\bar{\psi}i\gamma^\mu\mathcal{D}_\mu\psi\quad = \quad
\bar{\psi}i\gamma^\mu\left(\partial_\mu
- i g A_{\mu\alpha}T_\alpha\right)\psi\quad =
\qquad\qquad\qquad\qquad\qquad\qquad\qquad\qquad
\end{displaymath}
\begin{displaymath}
=\quad\bar{\psi}_{ijk}i\gamma^\mu\left(\partial_\mu
\frac{1}{6}\left(\delta_{mi}\delta_{pj}\delta_{qk}\pm\textrm{five 
terms}\right)\right.
\qquad\qquad\qquad\qquad\qquad\qquad\qquad\qquad
\end{displaymath}
\begin{displaymath}
\qquad\qquad\qquad\qquad
\left. - i g A_{\mu\alpha}\frac{1}{6}\left(-\left(t_\alpha
\right)_{mi}\delta_{pj}\delta_{qk}\pm\textrm{seventeen 
terms}\right)\right)\psi_{mpq} =
\end{displaymath}
\begin{equation}\label{massless Dirac action}
\qquad\qquad\qquad\qquad
=\quad\bar{\psi}_{ijk}i\gamma^\mu\partial_\mu\psi_{ijk}
- 3 g A_{\mu\alpha}\bar{\psi}_{ijk}\gamma^\mu \left(t_\alpha
\right)_{mi}\psi_{mjk}
\end{equation}
where I used (\ref{84 bar}), the antisymmetry of $\bar{\psi}_{ijk}$
and $\psi_{mpq}$ in their indexes, and the relabelling of dummy 
indexes.
$\bar{\psi}_{ijk}$ are the right-handed \textbf{84} states, and
$\psi_{mpq}$ are the left-handed $\mathbf{\overline{84}}$ states.

Breaking $\mathrm{SU}(9)$ to $\mathrm{SU}(3)\times\mathrm{SU}(2)
\times\mathrm{SU}(2)\times\mathrm{SU}(2)\times\mathrm{U}(1)_Y$, and
using the block matrix notation, this becomes:
\begin{displaymath}
\bar{\psi}_{BiCjDk}i\gamma^\mu\partial_\mu\psi_{BiCjDk}
- 3 g \sum_{A=1}^5 A_{\mu Aa}\bar{\psi}_{BiCjDk}\gamma^\mu 
\left(t_{Aa}^{(9)}\right)_{EmBi}\psi_{EmCjDk}\quad =\qquad\qquad
\qquad
\end{displaymath}
\begin{displaymath}
=\quad\left(\bar{\psi}_{BiCjDk}i\gamma^\mu\partial_\mu\psi_{BiCjDk}
- 3 g \sum_{A=1}^4 A_{\mu Aa}\bar{\psi}_{AiCjDk}\gamma^\mu 
\left(t_{Aa}\right)_{mi}\psi_{AmCjDk}\right.\quad\qquad\quad\quad
\end{displaymath}
\begin{equation}\label{after breaking}
\qquad\qquad\qquad\qquad\qquad\qquad\qquad\qquad\quad
\left. - 3 g A_{\mu 51} 
\frac{1}{\theta}
\displaystyle\sum_{A=1}^4\sigma_A
\bar{\psi}_{AiCjDk}\gamma^\mu\psi_{AiCjDk}\right)\quad
\end{equation}
where I used (\ref{non-Abelian generators}) and
(\ref{Abelian generator}).  We can now extract the covariant 
derivative Dirac action terms for the various entries in Table 
\ref{T3}, and thus determine the coefficients of their couplings to
$A_{\mu 51}$.  For example, a block in $\psi_{BiCjDk}$, where two
upper-case indexes take the value 1, and the remaining upper-case
index takes the value 2, 3, or 4, is a candidate to be a (3,2) quark
multiplet.  The sum of all terms in (\ref{after breaking}), where
two upper-case indexes take the value 1, and the remaining 
upper-case index takes the value 2, is:
\begin{displaymath}
\bigg(3\bar{\psi}_{1i1j2k}i\gamma^\mu\partial_\mu\psi_{1i1j2k}
- 6 g A_{\mu 1a}\bar{\psi}_{1i1j2k}\gamma^\mu\! 
\left(t_{1a}\right)_{mi}\psi_{1m1j2k}- 3 g A_{\mu 2a}\bar{\psi}_{2i1j1k}\gamma^\mu\! 
\left(t_{2a}\right)_{mi}\psi_{2m1j1k}
\end{displaymath}
\begin{equation}\label{112 after breaking}
\qquad\left.- 6 g A_{\mu 51} \frac{1}{\theta}\sigma_1
\bar{\psi}_{1i1j2k}\gamma^\mu\psi_{1i1j2k}
- 3 g A_{\mu 51} \frac{1}{\theta}\sigma_2
\bar{\psi}_{2i1j1k}\gamma^\mu\psi_{2i1j1k}\right)\quad
\end{equation}

Now $\psi_{1i1j2k}$ is antisymmetric under swapping the two
$\mathrm{SU}(3)$ antifundamental indexes $i$ and $j$, so that we may
write:
\begin{equation}\label{112 as quarks}
\psi_{1i1j2k}=\varepsilon_{ijm}\phi_{mk}
\end{equation}
and analogously:
\begin{equation}\label{RH 112 as antiquarks}
\bar{\psi}_{1i1j2k}=\varepsilon_{ijm}\bar{\phi}_{mk}
\end{equation}

We can then use relations such as
\begin{equation}\label{112 without inserted SU3 t}
\bar{\psi}_{1i1j2k}\gamma^\mu\psi_{1i1j2k}
=\varepsilon_{ijp}\bar{\phi}_{pk}\gamma^\mu
\varepsilon_{ijm}\phi_{mk}
=2\delta_{pm}\bar{\phi}_{pk}\gamma^\mu\phi_{mk}
=2\bar{\phi}_{mk}\gamma^\mu\phi_{mk}
\end{equation}
and
\begin{displaymath}
\bar{\psi}_{1i1j2k}\gamma^\mu\left(t_{1a}\right)_{mi}\psi_{1m1j2k}
=\varepsilon_{ijp}\bar{\phi}_{pk}\gamma^\mu\left(t_{1a}\right)_{mi}
\varepsilon_{mjq}\phi_{qk}=\qquad\qquad\qquad\qquad\qquad\qquad
\end{displaymath}
\begin{displaymath}
=\left(\delta_{im}\delta_{pq}-\delta_{iq}
\delta_{pm}\right)\bar{\phi}_{pk}\gamma^\mu\left(t_{1a}\right)_{mi}
\phi_{qk}=
\bar{\phi}_{pk}\gamma^\mu\left(t_{1a}\right)_{ii}\phi_{pk}-
\bar{\phi}_{mk}\gamma^\mu\left(t_{1a}\right)_{mi}\phi_{ik}=
\end{displaymath}
\begin{equation}\label{112 with inserted SU3 t}
\qquad\qquad\qquad\qquad\qquad\qquad\qquad\qquad\qquad
=-\bar{\phi}_{mk}\gamma^\mu\left(t_{1a}\right)_{mi}\phi_{ik}
\end{equation}
to express (\ref{112 after breaking}) as:
\begin{displaymath}
6\bigg(\bar{\phi}_{ij}i\gamma^\mu\partial_\mu\phi_{ij}
+ g A_{\mu 1a}
\bar{\phi}_{ij}\gamma^\mu\left(t_{1a}\right)_{ik}\phi_{kj}
- g A_{\mu 2a}\bar{\phi}_{ij}\gamma^\mu\! 
\left(t_{2a}\right)_{mj}\phi_{im}\qquad\qquad\qquad
\end{displaymath}
\begin{equation}\label{112 after breaking and transformation}
\qquad\qquad\qquad\qquad\qquad\qquad\qquad\qquad
\qquad\left.+ g A_{\mu 51} \frac{1}{\theta}\left( - 2 \sigma_1
- \sigma_2\right)\bar{\phi}_{ij}\gamma^\mu\phi_{ij}\right)\quad
\end{equation}

Thus we see that the index $i$ of $\phi_{ij}$ is an $\mathrm{SU}(3)$
fundamental index.  The $\mathrm{SU}(2)$ antifundamental is 
equivalent to the fundamental, the relation being given by matrix
multiplication by $\varepsilon_{jk}$, and we could, if we wished,
make a further transformation to replace the $\mathrm{SU}(2)$
antifundamental index $j$ of $\phi_{ij}$, by an index that is 
manifestly in the $\mathrm{SU}(2)$ fundamental.  When $(\mathrm{SU}
(2))^3$ is broken to $\mathrm{SU}(2)_{diag}$, the $A_{\mu2a}$, in
the third term in (\ref{112 after breaking and transformation}), 
will be replaced, at low energy, by $\frac{1}{\sqrt{3}}B_\mu$, where
$B_\mu$ is the gauge field of $\mathrm{SU}(2)_{diag}$.  The overall
factor of 6 can be absorbed into the normalizations of $\phi_{ij}$
and $\bar{\phi}_{ij}$, so from the fourth term in
(\ref{112 after breaking and transformation}), we can read off what 
the coefficient of $g A_{\mu 51}\bar{\phi}_{ij}\gamma^\mu\phi_{ij}$ 
would be, if the 
$\bar{\phi}_{ij}i\gamma^\mu\partial_\mu\phi_{ij}$ term had standard
normalization, and thus complete the entries in the second row of
Table \ref{T3}.

The entries in the third column of Tables \ref{T2} and \ref{T3} can
be completed by similar methods.  The entries in the fifth column 
of Table \ref{T3} 
can be completed by a simple mnemonic: for each upper-case index, of
the untransformed $\psi_{BiCjDk}$, that takes the value $A$, 
$1\leq A\leq 4$, include a term $-\frac{1}{\theta}\sigma_A$.  For
Table \ref{T2}, the mnemonic is that when the index $B$ of 
$\psi_{BiCj}$ takes the value $A$, $1\leq A\leq 4$, so that $i$ is 
in the fundamental of non-Abelian subgroup number $A$, include a
term $\frac{1}{\theta}\sigma_A$, and when the index $C$ of 
$\psi_{BiCj}$ takes the value $A$, $1\leq A\leq 4$, so that $j$ is 
in the antifundamental of non-Abelian subgroup number $A$, include 
a term $-\frac{1}{\theta}\sigma_A$.

Indeed, suppose we extract all terms from
(\ref{after breaking}) such that $\psi_{BiCjDk}$ has $n_A$ 
upper-case indexes with the value $A$, $1\leq A\leq 4$, so that
$n_1+n_2+n_3+n_4=3$.  We get $\frac{3!}{n_1!n_2!n_3!n_4!}$ 
contributions from the 
$\bar{\psi}_{BiCjDk}i\gamma^\mu\partial_\mu\psi_{BiCjDk}$ term.  The
number of times we get $\sigma_A$, from the third term in
(\ref{after breaking}), is 
$\frac{2!}{\tilde{n}_1!\tilde{n}_2!\tilde{n}_3!\tilde{n}_4!}$, where
$\tilde{n}_B=n_B$ if $B\neq A$, and $\tilde{n}_A=n_A-1$.  But this 
is equal to $\frac{2!n_A}{n_1!n_2!n_3!n_4!}$.  The factor
$\frac{2!}{n_1!n_2!n_3!n_4!}$ combines with the explicit factor of
3, in the third term in (\ref{after breaking}), to produce the same
overall factor of $\frac{3!}{n_1!n_2!n_3!n_4!}$ as found for the
first term, so the coefficient of the contributions from the third
term, if the contributions from the first term had standard
normalization, would be $-\frac{1}{\theta}\displaystyle\sum_{A=1}^4
n_A\sigma_A$.  The mnemonic for Table \ref{T2} can be justified in a
similar manner.

We know that we have to find couplings of the observed fermions, to
the $\mathrm{U}(1)_Y$ gauge field, that are smaller than those found
in the $\mathrm{SU}(5)$ model \cite{GG, Mohapatra}, by an overall
factor that is within a few percent of $\frac{1}{\sqrt{6}}$, so it 
is useful to apply the same techniques to calculate the 
corresponding coefficients in the $\mathrm{SU}(5)$ model.  In this
case, the relations (\ref{normalization}) and (\ref{tracelessness})
completely determine the $\mathrm{U}(1)$ generator, up to sign, and
we find the entries in the fourth column of Table \ref{T1}.  The
entries in the fifth column have been filled in, assuming the 
overall factor is exactly $\frac{1}{\sqrt{6}}$.

Comparing Tables \ref{T1}, \ref{T2}, and \ref{T3}, we see that for 
most of the entries in Table \ref{T1}, the left-handed \textbf{80} 
and $\mathbf{\overline{84}}$ can accomodate three left-handed
generations, with states to spare.  The three $\psi_\mathrm{diag}$
states, in Table \ref{T2}, are natural candidates for the 
left-handed antineutrinos, so a certain linear combination of the
left-handed antineutrinos, observed indirectly in oscillation
experiments, together with the corresponding linear combination of
the right-handed neutrinos, is the superpartner of the $U(1)_Y$ 
gauge field, or in other words, of $\cos\theta_WA_\mu-\sin\theta_W
Z_\mu$, \cite{Rosner},where $A_\mu$ denotes the photon, and $Z_\mu$
denotes the neutral weak vector boson.

For the $(\bar{3},1)$ left-handed antiquark states, we see that we
have exactly the right number of states to accomodate three 
generations, but there is a snag.  For both the $\psi_{221}$ family
and the $\psi_{123}$ family, the sum of the three entries, in the
fifth column of Table \ref{T3}, is equal to
\begin{equation}\label{sum of three entries}
-\frac{1}{\theta}
\left(3\sigma_1+2\sigma_2+2\sigma_3+2\sigma_4\right)
\end{equation}
which by (\ref{tracelessness}), is equal to zero.  Thus we cannot
accomodate three families of antiquarks whose weak hypercharges, 
$Y$, do not add up to zero.  The best we can do is accomodate two
up-type antiquarks, and four down-type antiquarks.  Thus the 
left-handed top antiquark cannot be accomodated, and must come, 
like the right-handed top quark, from
a non-zero Kaluza-Klein mode of the six-dimensional compact space 
that forms the inner surface of the cross-section of the pipe.  Thus
the top quark is the first observed state with a substantial
admixture, specifically 50\%, from a non-zero Kaluza-Klein mode.

The multiplets for which we have to find a $Y$-coefficient of the
smallest non-zero magnitude, specifically within a few percent of
$\frac{1}{\sqrt{360}}$, are the (3,2) quark multiplets, for which we
have six multiplets, (three in the \textbf{80}, and three in the
$\mathbf{\overline{84}}$), to accomodate the three observed 
multiplets.

Let us first try to find a one-to-one correspondence between three of the six available multiplets, and the three observed
multiplets.  There are then several cases to consider, depending on
which three of the six multiplets are assumed to be the observed 
ones.  In each case, requiring that the coefficient of the coupling
to the $\mathrm{U}(1)_Y$ gauge field be equal for the three observed
multiplets, together with the tracelessness condition 
(\ref{tracelessness}) and the normalization condition
(\ref{theta squared}), fixes the coefficient
uniquely, up to sign, and we find that the possible values of the 
coefficient are $\pm\frac{1}{2}$, $\pm\frac{1}{\sqrt{24}}$, and 
$\pm\frac{1}{\sqrt{132}}$.  None of these is within a few percent 
of $\frac{1}{\sqrt{360}}$, so we have 
exhausted the freedom to vary the embedding of $\mathrm{U}(1)_Y$,
without finding a solution.

There is another possibility, however.  We know that the masses of
the observed fermions break $\mathrm{SU}(2)\times\mathrm{U}(1)_Y$
invariance, conserving only the electromagnetic $\mathrm{U}(1)_Q$
subgroup, whose gauge field is the photon.  Furthermore, we expect,
from the existence of the CKM matrix, \cite{Cabibbo, KM, Rosner},
that the low-energy mass matrix, between the $\mathrm{E}8$ states 
of definite 
$\mathrm{SU}(3)\times\mathrm{SU}(2)\times\mathrm{U}(1)_Y$ quantum 
numbers, will be non-diagonal, and can include cross terms that violate both $\mathrm{SU}(2)$ conservation and $\mathrm{U}(1)_Y$
conservation, provided that $\mathrm{SU}(3)\times\mathrm{U}(1)_Q$ is
conserved.  In particular, the possibility that the low-energy mass
matrix includes cross terms between states of different $Y$, means
that for the (3,2) states, for example, the three low-energy mass
eigenstates can be three mutually orthogonal linear combinations of
the six available (3,2) states, and that each of these linear
combinations can include states with different $Y$ coefficients.

We can think of the sum of the couplings of the six (3,2) states, 
to the $\mathrm{U}(1)_Y$ gauge field, as a bilinear form between 
the six left-handed (3,2) states, and the six corresponding 
right-handed
$(\bar{3},2)$ states, which happens to be diagonal, in the sense 
that there are no cross terms between the different multiplets.  We
can represent this bilinear form schematically as
\begin{equation}\label{phi bar Y phi}
\bar{\phi}\;Y\phi
\end{equation}
where $Y$ is a diagonal six by six matrix, whose diagonal matrix
elements are the $Y$-coefficients of the six (3,2) states, read from
the fifth columns of Tables \ref{T2} and \ref{T3}.  Now suppose that
$\phi=U\chi$, and correspondingly, $\bar{\phi}=\bar{\chi}\;U^\dag$,
where the components of $\chi$ and $\bar{\chi}$ correspond to the
low-energy mass eigenstates, and $U$ is unitary.  Then
(\ref{phi bar Y phi}) is equal to 
\begin{equation}\label{chi bar U dag Y U chi}
\bar{\chi}\;U^\dag YU\chi
\end{equation}

In general, $\left(U^\dag YU\right)_{ij}$ will be non-diagonal.
However, so far only the matrix
elements between the lightest three of the six mass eigenstates,
which I shall take to be the states 1, 2, and 3, have been observed 
experimentally.  This 
submatrix is a multiple of the three by
three unit matrix, with the multiple being $\frac{1}{\sqrt{360}}$,
within a few percent.

The masses of the three heavy mass eigenstates, not yet observed, 
break $\mathrm{SU}(2)\times\mathrm{U}(1)_Y$ invariance, and thus 
can not be more than a few hundred GeV, \cite{Adler}, so these 
states should be produced copiously at the LHC.  In addition, the
off-diagonal matrix elements, in (\ref{chi bar U dag Y U chi}), 
between the three light mass eigenstates, and the three heavy mass
eigenstates, will result in new point-like contributions to 
processes such as $\bar{u}u\to\gamma\gamma$ and $\bar{u}u\to 
Z^0Z^0$, resulting from the exchange of one of the heavy fermions.

Using the unitarity of $U$, the low-energy mass eigenstates, 
$\chi_i$ and $\bar{\chi}_i$, are expressed in terms of $\phi_i$ and
$\bar{\phi}_i$ by:
\begin{equation}\label{chi in terms of phi}
\chi_i=\left(U_{ki}\right)^*\phi_k
\end{equation}
\begin{equation}\label{chi bar in terms of phi bar}
\bar{\chi}_i=\bar{\phi}_kU_{ki}
\end{equation}

Now, temporarily dropping the summation convention, if we write
$Y_{km}=y_k\delta_{km}$, where $y_k$ are the $Y$-coefficients of the
six (3,2) states, read from the fifth column of Tables \ref{T2} and
\ref{T3}, we can write:
\begin{equation}\label{U dag Y U in terms of the Y coefficients}
\left(U^\dag YU\right)_{ij}=\sum_{k=1}^6y_k\left(U_{ki}\right)^*
\left(U_{kj}\right)
\end{equation}

Finally, restoring the summation convention, the unitarity of $U$
can be expressed as:
\begin{equation}\label{unitarity of U}
\left(U_{ki}\right)^*U_{kj}=\delta_{ij}
\end{equation}

Looking at (\ref{chi in terms of phi}),
(\ref{chi bar in terms of phi bar}),
(\ref{U dag Y U in terms of the Y coefficients}), and
(\ref{unitarity of U}), we see that all the information about the
three observed mass eigenstates, and the matrix elements of weak
hypercharge between them, is contained in the six $Y$-coefficients,
$y_k$, read from the fifth column of Tables \ref{T2} and \ref{T3},
and the matrix elements $U_{ki}$, $1\leq k\leq 6$, $1\leq i\leq 3$,
and their complex conjugates.  We do not need to know the $U_{ki}$
for $4\leq i\leq 6$.  The matrix elements $\left(U_{ki}\right)^*$, 
$1\leq k\leq 6$, $1\leq i\leq 3$, can moreover be displayed 
conveniently, by writing out the relations 
(\ref{chi in terms of phi}) explicitly, for $1\leq i\leq 3$.

This approach can be adapted directly to the other $\mathrm{SU}(3)
\times\mathrm{SU}(2)$ multiplets, by making the appropriate 
replacements for the number of available multiplets, (here 6), and
the number of observed multiplets, (here 3).

The most stringent constraints are likely to come from the
left-handed $(\bar{3},1)$ antiquark multiplets, since the number of
available multiplets is 6, and the number of observed multiplets is
5, (since the left-handed top antiquark multiplet must come from a
non-zero Kaluza Klein state).  Let us try for a solution with
$\sigma_3 = \sigma_4$.  We then find, from (\ref{tracelessness}), 
that $\sigma_3 = \sigma_4 = -\left(\frac{3\sigma_1+2\sigma_2}{4}
\right)$, so that the $Y$-coefficients of the six left-handed
$(\bar{3},1)$ antiquark multiplets are as in Table \ref{T4}.

\begin{table}
\begin{center}
\begin{tabular}{|c|c|c|}\hline
State & $Y$ coefficient &
$\begin{array}{c}Y\textrm{ coefficient} \\
\textrm{when }\sigma_3=\sigma_4 \end{array}$ \\ 
\hline
$\psi_{122}$ & $\frac{-\sigma_1-2\sigma_2}{\theta}$ &
$-\frac{\sigma_1+2\sigma_2}{\theta}$ \\ 
$\psi_{133}$ & $\frac{-\sigma_1-2\sigma_3}{\theta}$ &
$\frac{\sigma_1+2\sigma_2}{2\theta}$ \\ 
$\psi_{144}$ & $\frac{-\sigma_1-2\sigma_4}{\theta}$ &
$\frac{\sigma_1+2\sigma_2}{2\theta}$ \\ 
$\psi_{123}$ & $\frac{-\sigma_1-\sigma_2-\sigma_3}{\theta}$ &
$-\frac{\sigma_1+2\sigma_2}{4\theta}$ \\ 
$\psi_{124}$ & $\frac{-\sigma_1-\sigma_2-\sigma_4}{\theta}$ &
$-\frac{\sigma_1+2\sigma_2}{4\theta}$ \\ 
$\psi_{134}$ & $\frac{-\sigma_1-\sigma_3-\sigma_4}{\theta}$ &
$\frac{\sigma_1+2\sigma_2}{2\theta}$ \\ \hline
\end{tabular}
\caption{\label{T4}
$Y$ coefficients of the left-handed $(\bar{3},1)$ antiquark 
multiplets}
\end{center}
\end{table}

We can identify $\psi_{133}$ and $\psi_{144}$ as up-type antiquarks,
with $Y$ coefficient $\frac{\sigma_1+2\sigma_2}{2\theta}$, and,
using equation (\ref{U dag Y U in terms of the Y coefficients}),
$\psi_{123}$, $\psi_{124}$, and $\frac{1}{\sqrt{2}}\left(\psi_{122}+
\psi_{134}\right)$, as down-type antiquarks, with $Y$ coefficient
$-\frac{\sigma_1+2\sigma_2}{4\theta}$.  Thus, from Table \ref{T1},
we require
\begin{equation}\label{condition on the sigmas for the antiquarks}
\frac{\sigma_1+2\sigma_2}{2\theta}=\frac{-4}{\sqrt{360}}
\end{equation}

With (\ref{theta squared}), this implies
\begin{equation}\label{quadratic equation for sigma 2}
44\sigma_2^2+44\sigma_1\sigma_2-13\sigma_1^2=0
\end{equation}

The two solutions may be chosen as in Table \ref{T5}.

\begin{table}
\begin{center}
\begin{tabular}{|c|c|c|c|c|c|}\hline
 & $\sigma_1$ & $\sigma_2$ & $\sigma_3$ & $\sigma_4$ & $\theta$ \\
\hline
Solution A & 2 &
$\begin{array}{c}-1-2\sqrt{\frac{6}{11}} \\
-2.4770979 \end{array}$ &
$\begin{array}{c}-1+\sqrt{\frac{6}{11}} \\
-0.2614511 \end{array}$ &
$\begin{array}{c}-1+\sqrt{\frac{6}{11}} \\
-0.2614511 \end{array}$ &
$\begin{array}{c}6\sqrt{\frac{15}{11}} \\
7.0064905 \end{array}$ \\ \hline
Solution B & $-2$ &
$\begin{array}{c}1-2\sqrt{\frac{6}{11}} \\
-0.4770979 \end{array}$ &
$\begin{array}{c}1+\sqrt{\frac{6}{11}} \\
1.7385489 \end{array}$ &
$\begin{array}{c}1+\sqrt{\frac{6}{11}} \\
1.7385489 \end{array}$ &
$\begin{array}{c}6\sqrt{\frac{15}{11}} \\
7.0064905 \end{array}$ \\ \hline
\end{tabular}
\caption{\label{T5}
The two solutions for the $\mathrm{U}(1)_Y$ generator found for the
$(\bar{3},1)$ LH antiquark multiplets, assuming 
$\sigma_3=\sigma_4$.}
\end{center}
\end{table}

Choosing one or the other of these two solutions for the
$\mathrm{U}(1)_Y$ generator, we can now seek solutions for three
generations of each of the other fermion multiplets in \mbox{Table 
\ref{T1}}.  We have already noted that the three 
$\psi_{\mathrm{diag}}$ states, in the \textbf{80}, are natural
candidates for the three left-handed antineutrinos.  For each of the
other four multiplets in \mbox{Table \ref{T1}}, there are at least 
twice as
many available multiplets, in the \textbf{80} and the 
$\mathbf{\overline{84}}$, as we require for the three observed
generations.  Therefore we can simplify the search for linear
combinations of the available multiplets, that have the correct
$Y$-coefficients, and no cross-terms between the observed fermions,
by assuming that each observed fermion state is a linear combination
of at most two of the available multiplets, and that each available
multiplet contributes to at most one observed fermion state.  There
are then automatically no cross terms between distinct observed
fermions $i$ and $j$ in equation
(\ref{U dag Y U in terms of the Y coefficients}), and if, for 
example, the two available multiplets contributing to the observed
fermion state $i$, are $a$ and $b$, then equation
(\ref{U dag Y U in terms of the Y coefficients}) expresses the 
$Y$ coefficient of the observed state, $y_i$, in terms of the $Y$
coefficients of the available multiplets $a$ and $b$, as
\begin{equation}\label{Y coefficient of a convenient observed state}
y_i=y_a\vert U_{ai}\vert^2+y_b\vert U_{bi}\vert^2
\end{equation}

Equation (\ref{unitarity of U}) reduces to:
\begin{equation}\label{unitarity for a convenient observed state}
\vert U_{ai}\vert^2+\vert U_{bi}\vert^2=1
\end{equation}

The problem of finding the observed fermion state $i$ thus reduces
to solving (\ref{Y coefficient of a convenient observed state}) and
(\ref{unitarity for a convenient observed state}), as simultaneous
linear equations for $\vert U_{ai}\vert^2$ and
$\vert U_{bi}\vert^2$, with $y_i$ being the required $Y$ 
coefficient, read from the fifth column of Table \ref{T1}, and $y_a$
and $y_b$ being the $Y$ coefficients of the available multiplets $a$
and $b$, read from the fifth column of Table \ref{T2} or Table
\ref{T3}, after substituting in the parameters of the chosen 
solution A or B for the $\mathrm{U}(1)_Y$ generator, from Table 
\ref{T5}.  The solution of
(\ref{Y coefficient of a convenient observed state}) and
(\ref{unitarity for a convenient observed state}) is
\begin{equation}\label{solution of the two equations}
\vert U_{ai}\vert^2=\frac{y_i-y_b}{y_a-y_b}\:\:,
\qquad\qquad
\vert U_{bi}\vert^2=\frac{y_a-y_i}{y_a-y_b}
\end{equation}
so we find a fermion state $i$, with $\vert U_{ai}\vert^2\geq 0$ and
$\vert U_{bi}\vert^2\geq 0$, if, and only if,
$y_i$ lies between $y_a$ and $y_b$.  Thus we can obtain a solution
of this type, provided we can assign the observed fermion
states to mutually disjoint pairs of available multiplets,
such that the $Y$ coefficient of each observed fermion state lies
between the $Y$ coefficients of the pair of available multiplets it
is assigned to.

Let us try for a solution using solution A from Table \ref{T5}.  
Then for the (3,2) quark multiplets, we find that the $Y$ 
coefficients of $\psi_{211}$, $\psi_{311}$, and $\psi_{411}$ are
negative, and the $Y$ coefficients of $\psi_{12}$, $\psi_{13}$, and
$\psi_{14}$ are positive, and greater than the required value,
$\frac{1}{\sqrt{360}}$, so a possible solution is 
($\psi_{211}$,$\psi_{12}$), ($\psi_{311}$,$\psi_{13}$), 
($\psi_{411}$,$\psi_{14}$).
For the (1,2) lepton doublets, we find that all eight (1,2)
multiplets in the $\mathbf{\overline{84}}$ have positive $Y$
coefficients, whereas the required $Y$ coefficient,
$\frac{-3}{\sqrt{360}}$, is negative.  However, since the 
$\mathrm{SU}(2)$ fundamental is equivalent to the antifundamental,
the relation being given by matrix multiplication by 
$\varepsilon_{ij}$, we can also use the (1,2) multiplets in the
left-handed \textbf{84}, which have negative $Y$ coefficients.  Six
of the (1,2) multiplets in the left-handed \textbf{84} have 
negative $Y$ coefficients of larger magnitude than the required
value, and two have negative $Y$ coefficients of smaller magnitude
than the required value, so we can obtain a solution, for example,
with two pairs of (1,2) multiplets coming from the \textbf{84}, and
one pair of (1,2) multiplets with one member from the \textbf{84},
and one member from the $\mathbf{\overline{84}}$.  For the $e^+$,
$\mu^+$, and $\tau^+$ (1,1) states, we find that $\psi_{32}$ and
$\psi_{42}$ have exactly the required $Y$ coefficient, 
$\frac{6}{\sqrt{360}}$, and thus may be identified with two of these
states.  The third state may be obtained as a linear combination of
$\psi_{111}$ in the \textbf{84}, and, for example, $\psi_{23}$ or
$\psi_{24}$.  Finally, we note that $\psi_{34}$ and $\psi_{43}$ have
zero $Y$ coefficient, and can mix with the three 
$\psi_{\mathrm{diag}}$ antineutrino states.

Once the observed fermion states with a given $\mathrm{SU}(3)\times
\mathrm{SU}(2)$ content, and a given $Y$ coefficient, have been
obtained, they can be mixed among themselves without producing
cross-terms.  A further mixing possibility, without producing
cross-terms, is to mix $\mathrm{E}8$ states that have the same
$\mathrm{SU}(3)\times\mathrm{SU}(2)$ content and $Y$ coefficients,
among themselves, before assigning them to mutally disjoint pairs
associated with the observed fermions.

The stability of the proton, in this class of models, must be a
dynamical property of the vacuum, which is not yet understood.  Thus
the natural way to attempt to fit the observed stability of the
proton, would be to attempt to adjust the substantial number of
mixing angles, relating the observed fermion states to the 
$\mathrm{E}8$ states, so as to cancel the most dangerous proton
decay modes.  Some of these mixing angles will affect the CKM 
matrix, so this suggests that the entries in the CKM matrix might be
correlated with the stability of the proton.

\end{document}